\documentclass[11pt]{article}
\begin{document}


\begin{flushright}
   MZ-TH--97-02  \\
   hep-ph/9701422 \\
   January 1997 \\
\end{flushright}
\vspace*{3cm}
\begin{center} {\bf \large 
  Corrections to Sirlin's Theorem 
  in ${\cal O}(p^6)$ Chiral Perturbation Theory } 
  \vspace{10 mm} \\
  P.~Post and K.~Schilcher
  \footnote{E-mail: post@vipmzw.physik.uni-mainz.de,
               schilcher@vipmzw.physik.uni-mainz.de}
  \vspace{10 mm} \\
  {\em Institut f\"ur Physik, Johannes Gutenberg Universit\"at,\\
  Staudinger Weg 7, D-55099 Mainz, Germany}
\end{center}
\vspace{12 mm}
\begin{abstract}
We present the results of the first two-loop calculation of a form factor
in full $SU(3) \times SU(3)$ Chiral Perturbation Theory. We choose a
specific linear combination of $\pi^+$, $K^+$, $K^0$ and $K\pi$ form
factors (the one appearing in Sirlin's theorem) which does not get
contributions from order $p^6$ operators with unknown constants.  For the
charge radii, the correction to the previous one-loop result turns out to
be significant, but still there is no agreement with the present data due
to large experimental uncertainties in the kaon charge radii.
\end{abstract}

\newpage

Chiral perturbation theory (ChPT) has been applied with great success to
low energy hadronic phenomena \cite{Weinberg,GasserLeutw1,GasserLeutw2}.
Presently there is an emerging efford to extend the calculations to
two-loop order so as to allow quantitative comparison with experiments and
test the convergence properties of ChPT. Up to now most of the two-loops
results have been obtained in the chiral $SU(2)\times SU(2)$-limit
which is obviously a serious limitation as $K$-mesons are excluded from
loops. To our knowledge a complete $SU(3)\times SU(3)$ calculation
exists so far only for two-point-functions of current-correlators
\cite{KamborGolo}. In this note we will present the results of the first
full $SU(3)\times SU(3)$ form factor calculation. We choose a specific
combination of weak and electromagnetic meson form factors which does not
involve arbitrary renormalization constants of new operators.

ChPT is formulated in terms of an effective Lagrangian involving an 
increasing number of covariant derivatives, external fields (including 
quark mass terms) and field strength tensors,
\begin{eqnarray}
{\cal L}_{\mbox{\scriptsize eff}}
&=& {\cal L}^{(2)} + {\cal L}^{(4)} + {\cal L}^{(6)} + \ldots \:.
\end{eqnarray}  
The lowest-order term is \cite{Weinberg}
\begin{eqnarray}
{\cal L}^{(2)} &=&
\frac{F^2}{4} \: \mbox{\rm Tr}\big( D_\mu\!\!\;U D^\mu\!\!\:U^\dagger \big)
+ \frac{F^2}{4} \: \mbox{\rm Tr}\big( \chi U^\dagger + U \chi^\dagger \big) 
\end{eqnarray}
with $U(x)=\exp[i \Phi(x)/F]$ where $\Phi$ is the $3\times 3$ 
matrix made up of the Goldstone fields $(\pi,K,\eta)$, $F$ is the pion 
decay constant in the chiral limit, and $\chi$ is related to the quark mass 
matrix (for details see \cite{GasserLeutw1}). 
The term ${\cal L}^{(4)}$ is of order $p^4$ and involves 10 new operators
which are to be renormalized by imposing the same number of independent
experimental input data. The renormalization constants $L_1$ to $L_{10}$
are commonly defined in dimensional regularization. The operators of
${\cal L}^{(6)}$ have been exhaustively analysed in \cite{FearingScherer}
and found to be 143 in number. The corresponding number of free constants
seems to be prohibitive. There are, however, subsets of experiments like
the weak and electromagnetic form factors of mesons which involve only a
small number of new renormalization constants. ChPT to order $p^6$ then
leads to relations between and predictions of specifics of their
$t$-dependence. In one combination of these form factors the ${\cal
L}^{(6)}$-constants all cancel. This is the combination entering Sirlin's
relation \cite{Sirlin} and which should vanish in the chiral limit.

The relevant vector current form factors are defined as follows:
\begin{eqnarray}
\langle \pi^+,p^\prime \mid J_\mu \mid \pi^+,p \rangle &=& 
(p+p^\prime)_\mu \, F^\pi(t)
\\
\langle K,p^\prime \mid J_\mu \mid K,p \rangle 
&=& (p+p^\prime)_\mu \, F^K(t)
\\
\langle \pi^-,p^\prime \mid \bar{u} \gamma_\mu s \mid K^0,p \rangle 
&=& (p+p^\prime)_\mu \, f_+^{K\pi}(t) 
+ (p-p^\prime)_\mu \, f_-^{K\pi}(t) \:,
\end{eqnarray}
where $t=(p^\prime-p)^2$ and
$J_\mu$ is the electromagnetic current carried by the light quarks,
$J_\mu=\frac{2}{3} \bar{u}\gamma_\mu u 
- \frac{1}{3} \bar{d}\gamma_\mu d
- \frac{1}{3} \bar{s}\gamma_\mu s$.
Sirlin's low-energy theorem then states that up to second order in the
quark mass difference $m_s-\hat{m}$, 
$\hat{m}=\frac{1}{2} (m_u+m_d)$, the linear combination
\begin{eqnarray}\label{Sirlins_relation}
\Delta(t) &=& \frac{1}{2} F^{\pi^+}(t) + \frac{1}{2} F^{K^+}(t) 
+ F^{K^0}(t) - f_+^{K\pi}(t) 
\end{eqnarray}
vanishes. The effect of heavy quarks in the electromagnetic current is
neglected. Sirlin's relation generalizes the Ademollo Gatto theorem
\cite{AdemolloGatto} to $t\neq 0$. 
For $t=0$ eqn.$\,$(\ref{Sirlins_relation}) yields no prediction as
$f^{K\pi}_+(0)$ still depends on unknown constants of ${\cal L}^{(6)}$. 
In the relation for the charge radii (and higher Taylor coefficients),
however, all arbitrary constants cancel and an unambiguous prediction
remains.

\begin{figure}[t] 
\begin{center}
\unitlength0.7mm
\begin{picture}(165,35) 
\put(0,30){\begin{picture}(0,0)\thicklines\unitlength0.7mm  
  \put(15,0){\makebox(0,0)[b]{diagram (a)}}
  \put(15,-5){\begin{picture}(0,0)\thicklines\unitlength0.7mm
              \put(0,0){\line(-1,0){15}} 
              \put(0,0){\line(1,0){15}}
              \put(0,0){\circle*{2}}
              \put(0,-7){\circle{14}}
              \put(0,-14){\circle*{2}}
              \put(0,-15){\begin{picture}(0,0) 
                 \put(0,-1){\oval(2.1,2.1)[r]}
                 \put(0,-3){\oval(2.1,2.1)[l]}
                 \put(0,-5){\oval(2.1,2.1)[r]}  
                 \put(0,-7){\oval(2.1,2.1)[l]}
                 \put(0,-9){\oval(2.1,2.1)[r]}
                 \end{picture}}
              \end{picture}}
  \put(60,0){\makebox(0,0)[b]{diagram (b)}}
  \put(60,-5){\begin{picture}(0,0)\thicklines\unitlength0.7mm
              \put(0,0){\line(-1,0){15}}
              \put(0,0){\line(1,0){15}}
              \put(0,0){\makebox(0,0){\rule[-0.75mm]{1.5mm}{1.5mm}}}
              \put(0,-1){\begin{picture}(0,0) 
                 \put(0,-1){\oval(2.1,2.1)[r]}
                 \put(0,-3){\oval(2.1,2.1)[l]}
                 \put(0,-5){\oval(2.1,2.1)[r]}  
                 \put(0,-7){\oval(2.1,2.1)[l]}
                 \put(0,-9){\oval(2.1,2.1)[r]}
                 \end{picture}}
              \end{picture}}
  \put(105,0){\makebox(0,0)[b]{diagram (c)}}
  \put(105,-5){\begin{picture}(0,0)\thicklines\unitlength0.7mm
               \put(0,0){\line(-1,0){15}}
               \put(0,0){\line(1,0){15}}
               \put(0,0){\makebox(0,0){\rule[-0.75mm]{1.5mm}{1.5mm}}}
               \put(0,-7){\circle{14}}
               \put(0,-14){\circle*{2}}
               \put(0,-15){\begin{picture}(0,0) 
                  \put(0,-1){\oval(2.1,2.1)[r]}
                  \put(0,-3){\oval(2.1,2.1)[l]}
                  \put(0,-5){\oval(2.1,2.1)[r]}  
                  \put(0,-7){\oval(2.1,2.1)[l]}
                  \put(0,-9){\oval(2.1,2.1)[r]}
                  \end{picture}}
               \end{picture}}
  \put(150,0){\makebox(0,0)[b]{diagram (d)}}
  \put(150,-5){\begin{picture}(0,0)\thicklines\unitlength0.7mm
               \put(0,0){\line(-1,0){15}}
               \put(0,0){\line(1,0){15}}
               \put(0,0){\circle*{2}}
               \put(0,-7){\circle{14}}
               \put(0,-14){\circle*{2}}
               \put(7,-7){\makebox(0,0){\rule[-0.75mm]{1.5mm}{1.5mm}}} 
               \put(0,-15){\begin{picture}(0,0) 
                  \put(0,-1){\oval(2.1,2.1)[r]}
                  \put(0,-3){\oval(2.1,2.1)[l]}
                  \put(0,-5){\oval(2.1,2.1)[r]}  
                  \put(0,-7){\oval(2.1,2.1)[l]}
                  \put(0,-9){\oval(2.1,2.1)[r]}
                  \end{picture}}
               \end{picture}}
  \end{picture}}
\end{picture}
\end{center}
\begin{center}
\unitlength0.7mm
\begin{picture}(165,40)
\put(0,39){\begin{picture}(0,0)\thicklines\unitlength0.7mm  
  \put(15,0){\makebox(0,0)[b]{diagram (e)}}
  \put(15,-5){\begin{picture}(0,0)\thicklines\unitlength0.7mm
              \put(0,0){\line(-1,0){15}}
              \put(0,0){\line(1,0){15}}
              \put(0,0){\circle*{2}}
              \put(0,-7){\circle{14}}
              \put(0,-14){\makebox(0,0){\rule[-0.75mm]{1.5mm}{1.5mm}}}
              \put(0,-15){\begin{picture}(0,0) 
                 \put(0,-1){\oval(2.1,2.1)[r]}
                 \put(0,-3){\oval(2.1,2.1)[l]}
                 \put(0,-5){\oval(2.1,2.1)[r]}  
                 \put(0,-7){\oval(2.1,2.1)[l]}
                 \put(0,-9){\oval(2.1,2.1)[r]}
                 \end{picture}}
              \end{picture}}
  \put(60,0){\makebox(0,0)[b]{diagram (f)}}
  \put(60,-17){\begin{picture}(0,0)\thicklines\unitlength0.7mm
               \put(0,0){\line(-1,0){15}}
               \put(0,0){\line(1,0){15}}
               \put(0,0){\makebox(0,0){\rule[-0.75mm]{1.5mm}{1.5mm}}} 
               \put(0,7){\circle{14}}
               \put(0,-1){\begin{picture}(0,0) 
                  \put(0,-1){\oval(2.1,2.1)[r]}
                  \put(0,-3){\oval(2.1,2.1)[l]}
                  \put(0,-5){\oval(2.1,2.1)[r]}  
                  \put(0,-7){\oval(2.1,2.1)[l]}
                  \put(0,-9){\oval(2.1,2.1)[r]}
                  \end{picture}}
               \end{picture}}
  \put(105,0){\makebox(0,0)[b]{diagram (g)}}
  \put(105,-5){\begin{picture}(0,0)\thicklines\unitlength0.7mm
               \put(0,0){\line(-1,0){15}}
               \put(0,0){\line(1,0){15}}
               \put(0,0){\circle*{2}}
               \put(0,-7){\circle{14}}
               \put(0,-14){\circle*{2}}
               \put(0,-21){\circle{14}}
               \put(0,-28){\circle*{2}}
               \put(0,-29){\begin{picture}(0,0) 
                 \put(0,-1){\oval(2.1,2.1)[r]}
                 \put(0,-3){\oval(2.1,2.1)[l]}
                 \put(0,-5){\oval(2.1,2.1)[r]}  
                 \put(0,-7){\oval(2.1,2.1)[l]}
                 \put(0,-9){\oval(2.1,2.1)[r]}
                 \end{picture}}
               \end{picture}}
  \put(150,0){\makebox(0,0)[b]{diagram (h)}}
  \put(150,-5){\begin{picture}(0,0)\thicklines\unitlength0.7mm
               \put(0,0){\line(-1,0){15}}
               \put(0,0){\line(1,0){15}}
               \put(0,0){\circle*{2}}
               \put(0,-7){\circle{14}}
               \put(0,-14){\circle*{2}}
	       \put(0,-21){\circle{14}}
               \put(0,-15){\begin{picture}(0,0) 
                  \put(0,-1){\oval(2.1,2.1)[r]}
                  \put(0,-3){\oval(2.1,2.1)[l]}
                  \put(0,-5){\oval(2.1,2.1)[r]}  
                  \put(0,-7){\oval(2.1,2.1)[l]}
                  \put(0,-9){\oval(2.1,2.1)[r]}
                  \end{picture}}
               \end{picture}}
  \end{picture}}
\end{picture}
\end{center}
\begin{center}
\unitlength0.7mm
\begin{picture}(165,42)
\put(0,37){\begin{picture}(0,0)\thicklines\unitlength0.7mm  
  \put(15,0){\makebox(0,0)[b]{diagram (i)}}
  \put(15,-5){\begin{picture}(0,0)\thicklines\unitlength0.7mm
              \put(0,0){\line(-1,0){15}}
              \put(0,0){\line(1,0){15}}
              \put(0,0){\circle*{2}}
              \put(0,-7){\circle{14}}
              \put(0,-14){\circle*{2}}
              \put(0,-21){\circle{14}}
              \put(7,-7){\circle*{2}}
              \put(7.5,-7){\begin{picture}(0,0) 
                 \put(1,0){\oval(2.1,2.1)[b]}  
                 \put(3,0){\oval(2.1,2.1)[t]}
                 \put(5,0){\oval(2.1,2.1)[b]}
                 \put(7,0){\oval(2.1,2.1)[t]}
                 \put(9,0){\oval(2.1,2.1)[b]}
		 \end{picture}}
              \end{picture}}
  \put(60,0){\makebox(0,0)[b]{diagram (j)}}
  \put(60,-17){\begin{picture}(0,0)\thicklines\unitlength0.7mm
               \put(0,0){\line(-1,0){15}}
               \put(0,0){\line(1,0){15}}
               \put(0,0){\circle*{2}}
               \put(0,7){\circle{14}}
               \put(0,-7){\circle{14}}
               \put(0,-14){\circle*{2}}
               \put(0,-15){\begin{picture}(0,0) 
                  \put(0,-1){\oval(2.1,2.1)[r]}
                  \put(0,-3){\oval(2.1,2.1)[l]}
                  \put(0,-5){\oval(2.1,2.1)[r]}  
                  \put(0,-7){\oval(2.1,2.1)[l]}
                  \put(0,-9){\oval(2.1,2.1)[r]}
                  \end{picture}}
               \end{picture}}
  \put(105,0){\makebox(0,0)[b]{diagram (k)}}
  \put(105,-12){\begin{picture}(0,0)\thicklines\unitlength0.7mm
                \put(0,0){\oval(17,15)}
                \put(0,0){\line(-1,0){18}}
                \put(0,0){\line(1,0){18}}
                \put(-8.5,0){\circle*{2}}
                \put(8.5,0){\circle*{2}}
                \put(0,-7.5){\circle*{2}}
                \put(0,-8.5){\begin{picture}(0,0) 
                   \put(0,-1){\oval(2.1,2.1)[r]}
                   \put(0,-3){\oval(2.1,2.1)[l]}
                   \put(0,-5){\oval(2.1,2.1)[r]}  
                   \put(0,-7){\oval(2.1,2.1)[l]}
                   \put(0,-9){\oval(2.1,2.1)[r]}
                   \end{picture}}
                \end{picture}}
  \put(150,0){\makebox(0,0)[b]{diagram (l)}}
  \put(150,-12){\begin{picture}(0,0)\thicklines\unitlength0.7mm
                \put(-1,1){\line(1,0){2}}
                \put(-1,1){\line(0,-1){2}}
                \put(1,1){\line(0,-1){2}}
                \put(-1,-1){\line(1,0){2}}
                \put(-1,0){\line(-1,0){14}}
                \put(1,0){\line(1,0){14}}
                \put(0,-1){\begin{picture}(0,0) 
                  \put(0,-1){\oval(2.1,2.1)[r]}
                  \put(0,-3){\oval(2.1,2.1)[l]}
                  \put(0,-5){\oval(2.1,2.1)[r]}  
                  \put(0,-7){\oval(2.1,2.1)[l]}
                  \put(0,-9){\oval(2.1,2.1)[r]}
                  \end{picture}}
                \end{picture}}
  \end{picture}}
\end{picture}
\end{center}
\normalsize Figure 1: 
\footnotesize 
The form factor diagrams with $t$-dependence. 
${\cal L}^{(2)}$-vertices are denoted by filled circles
\parbox{2.1mm}{\unitlength0.7mm\begin{picture}(0,0) 
               \put(1.5,0){\circle*{2}} 
               \end{picture}},   
${\cal L}^{(4)}$-vertices by filled squares 
\parbox{2.25mm}{\unitlength0.75mm\begin{picture}(0,0) 
            \put(1.5,0){\makebox(0,0){\rule[-0.666mm]{1.333mm}{1.333mm}}}
            \end{picture}}
and an ${\cal L}^{(6)}$-vertex by an open square
\parbox{2.25mm}{\unitlength0.75mm\begin{picture}(0,0) 
              \put(1.5,0){\begin{picture}(0,0) 
                 \put(-1,1){\line(1,0){2}}
                 \put(-1,1){\line(0,-1){2}}
                 \put(1,1){\line(0,-1){2}}
                 \put(-1,-1){\line(1,0){2}}
                 \end{picture}}
              \end{picture}}. 
\end{figure} 

The diagrams contributing to the $t$-dependence of the form factors
at order $p^6$ are represented symbollically in fig.$\,$(1). 
There are further graphs which do not depend on $t$ and are omitted here.
The ${\cal O}(p^6)$ contributions arise from the lowest order diagrams 
(a),(b) owing to wave function, mass and decay constant renormalization, 
from the one-loop-diagrams (c)--(f) with one vertex from ${\cal L}^{(4)}$,
from the reducible two-loop diagrams (g)--(j), from the irreducible
two-loop diagram (k) and from the tree graph (l) with one vertex from
${\cal L}^{(6)}$, which yields no contributions to Sirlin's linear
combination $\Delta(t)$. Diagrams (b),(f) and (l) are polynomial in $t$ 
due to the derivative couplings in the vertices.

All reducible diagrams (a)--(j) involve only well-known one-loop integrals
\cite{tHofftVeltmann} (calculated to order $\varepsilon=2-D/2$), 
i.e. the massive one-loop tadpole
\begin{eqnarray}
A(m^2) &=& 
\mu^{4-D} \int \frac{d^D\!\!\:k}{i(2\pi)^D} \; \frac{1}{k^2-m^2} 
\end{eqnarray}
and the one-loop two-point-functions $B_{21}(q^2;m_1^2,m_2^2)$
and $B_{31}(q^2;m_1^2;m_2^2)$ defined by the tensor decompositions
\begin{eqnarray}
\label{KontrGlBFkt}
\mu^{4-D} \int \frac{d^D\!\!\:k}{i(2\pi)^D} \; 
\frac{k^{\mu} \, k^{\nu}}{[(k+q)^2-m_1^2]\:[k^2-m_2^2]} 
&=:& q^\mu q^\nu \: B_{20}(q^2;m_1^2,m_2^2) 
+ g^{\mu\nu} \: B_{21}(q^2;m_1^2,m_2^2)
\\
\mu^{4-D} \int \frac{d^D\!\!\:k}{i(2\pi)^D} \; 
\frac{k^{\mu} \, k^{\nu} \, k^\rho}{[(k+q)^2-m_1^2]\:[k^2-m_2^2]} 
&=:& q^\mu  q^\nu  q^\rho \: B_{30} 
+ (g^{\mu\nu} q^\rho + g^{\mu\rho} q^\nu + g^{\nu\rho} q^\mu) \: B_{31} \:.
\quad
\end{eqnarray}
In terms of these integrals Sirlin's relation reads
\begin{eqnarray}
\Delta(t) &=& \delta_4(t) + \delta_6(t) + \ldots
\end{eqnarray}
where $\delta_4$ is the ${\cal O}(p^4)$-result \cite{GasserLeutw2},
\begin{eqnarray}
\delta_4(t) &=& 
\frac{1}{8 F^2} \Big\{
-3 \:\! A(m_\eta^2)
+4 \:\! A(m_K^2)
-A(m_\pi^2)
+12 \:\! B_{21}(q^2;m_\eta^2,m_K^2)
\\ && \quad
- \; 20 \: B_{21}(q^2;m_K^2,m_K^2)
+12 \: B_{21}(q^2;m_K^2,m_\pi^2)
-4 \: B_{21}(q^2;m_\pi^2,m_\pi^2)
\Big\} \:, \nonumber
\end{eqnarray}
and $\delta_6$ the ${\cal O}(p^6)$-contribution. Up to a constant which
is irrelevant for the $t$-dependence of $\Delta(t)$, $\delta_6$ is given
as a sum 
\begin{eqnarray}
\delta_6(t) &=&
\mbox{\em Red$_1(t)$} + \mbox{\em Red$_2(t)$} + \mbox{\em Irr$(t)$} 
+ \mbox{\em const} \:,
\end{eqnarray}
where the reducible one-loop part \mbox{\em Red$_1(t)$} collects all terms 
involving ${\cal L}^{(4)}$-parameters $L_1,\ldots,L_{10}$,
\begin{eqnarray}
\mbox{\em Red}_1(t) &=& \frac{1}{4 F^4} \Big\{
-3 \: A(m_\eta^2) \: q^2 L_9
+4 \: A(m_K^2) \: q^2 L_9
-A(m_\pi^2) \: q^2 L_9
\\&& \quad
+ \; 16 \: B_{31}(q^2;m_\eta^2;m_K^2) \: L_3 (m_K^2 - m_\pi^2)  
\nonumber\\&& \quad
- \; 16 \: B_{31}(q^2;m_K^2;m_\pi^2)
\: (8 L_1 + 4 L_2 + L_3) (m_K^2 - m_\pi^2)
\nonumber\\&& \quad
+ \; 4 \: B_{21}(q^2;m_\eta^2,m_K^2) \: [ 
2 L_3 (m_K^2 - m_\pi^2 - 3 q^2) + 12 L_5 m_\pi^2 + 3 q^2 L_9 ]
\nonumber\\&& \quad
+ \; 4 \: B_{21}(q^2;m_K^2,m_K^2) \: [ 
8 q^2 L_1 
-4 q^2 L_2
+10 q^2 L_3
-16 L_4 m_K^2
-20 L_5 m_\pi^2
-5 q^2 L_9 ]
\nonumber\\&& \quad
+ \; 4 \: B_{21}(q^2;m_\pi^2,m_\pi^2) \: [ 8 q^2 L_1 
-4 q^2 L_2
+2 q^2 L_3
-16 L_4 m_\pi^2
-4 L_5 m_\pi^2
-q^2 L_9 ]
\nonumber\\&& \quad
+ \; 4 \: B_{21}(q^2;m_K^2,m_\pi^2) \: [
-16 L_1 (m_K^2 - m_\pi^2 + q^2)
-8 L_2 (m_K^2 - m_\pi^2 - q^2) 
\nonumber\\&& \hspace{7em}
-2 L_3 (m_K^2 - m_\pi^2 + 3 q^2)
+16 L_4 (m_K^2 + m_\pi^2)
+12 L_5 m_\pi^2
+3 q^2 L_9 ] \Big\} \:, \nonumber
\end{eqnarray}
\mbox{\em Red$_2(t)$} denotes the reducible two-loop parts,
\begin{eqnarray}
\mbox{\em Red}_2(t) &=& \frac{1}{144 \: F^4} \Big\{
 99 \: A(m_\eta^2) \: B_{21}(q^2;m_\eta^2,m_K^2)
+102 \: A(m_\eta^2) \: B_{21}(q^2;m_K^2,m_K^2)
\\ && \quad
+ \; 75 \: A(m_\eta^2) \: B_{21}(q^2;m_K^2,m_\pi^2)
-6 \: A(m_\eta^2) \: B_{21}(q^2;m_\pi^2,m_\pi^2)
\nonumber \\ && \quad
+ \; 318 \: A(m_K^2) \: B_{21}(q^2;m_\eta^2,m_K^2)
-924 \: A(m_K^2) \: B_{21}(q^2;m_K^2,m_K^2)
\nonumber \\ && \quad
+ \; 350 \: A(m_K^2) \: B_{21}(q^2;m_K^2,m_\pi^2)
-104 \: A(m_K^2) \: B_{21}(q^2;m_\pi^2,m_\pi^2)
\nonumber \\ && \quad
+ \; 483 \: A(m_\pi^2) \: B_{21}(q^2;m_\eta^2,m_K^2)
-678 \: A(m_\pi^2) \: B_{21}(q^2;m_K^2,m_K^2)
\nonumber \\ && \quad
+ \; 475 \: A(m_\pi^2) \: B_{21}(q^2;m_K^2,m_\pi^2)
-190 \: A(m_\pi^2) \: B_{21}(q^2;m_\pi^2,m_\pi^2)
\nonumber \\ && \quad
- \; 324 \: B_{21}(q^2;m_\eta^2,m_K^2)^2
+1008 \: B_{21}(q^2;m_K^2,m_K^2)^2
\nonumber \\ && \quad
+ \; 144 \: B_{21}(q^2;m_\pi^2,m_\pi^2)^2
-648 \: B_{21}(q^2;m_\eta^2,m_K^2) 
     \: B_{21}(q^2;m_K^2,m_\pi^2)
\nonumber \\ && \quad
- \; 324 \: B_{21}(q^2;m_K^2,m_\pi^2)^2
+144 \: B_{21}(q^2;m_K^2,m_K^2) 
     \: B_{21}(q^2;m_\pi^2,m_\pi^2)
\Big\}  \nonumber
\end{eqnarray}
and \mbox{\em Irr(t)} the irreducible two-loop contribution from diagram (k).
Diagram (k) cannot be calculated analytically, unless all masses are
equal. Instead, for arbitrary masses and tensor numerators, it can be
reduced via dispersion techniques to a one-dimensional integral which is
done numerically \cite{PostTausk,PostSchilcher}. 
Its contribution to the charge radius
\begin{equation}
\langle r^2 \rangle_{\mbox{\em\scriptsize Sirlin}}
\;:=\; 6\, \frac{d\Delta}{dt}\Big|_{t=0}
\end{equation}  
turns out to be small if one uses the generalized Gasser-Leutwyler
renormalization scheme, i.e. multiplication of each 
${\cal O}(p^6)$-contribution with the $\overline{\mbox{MS}}$-type factor
\begin{equation}
\Big[ \varepsilon \!\: (4\pi)^\varepsilon \!\; 
\Gamma(-1+\varepsilon) \Big]^2
\;=\;
\exp\Big[ 2 \!\: \varepsilon \!\: (\gamma-1-\log 4\pi)
          -\varepsilon^2 \Big( \frac{\pi^2}{6} + 1 \Big)
          +{\cal O}(\varepsilon^3) \Big] \:.
\end{equation}
For $\mu=770 \:\mbox{MeV}$, $F_\pi=92.4\:\mbox{MeV}$, 
$m_K=495\:\mbox{MeV}$, $m_\pi=135\:\mbox{MeV}$ 
and $m_\eta=548.8\:\mbox{MeV}$ we find
\begin{eqnarray}
\langle r^2 \rangle_{\mbox{\em\scriptsize Sirlin}}
&=& \big[ 0.006 \: \mbox{fm}^2 \big]^{{\cal O}(p^4)}
             \;+\; \big[ 0.017(3) \: \mbox{fm}^2 \big]^{
                    \mbox{\em\scriptsize red.$\;\!{\cal O}(p^6)$}}
             \;+\; \big[ -0.002 \: \mbox{fm}^2 \big]^{
                    \mbox{\em\scriptsize irr.$\;\!{\cal O}(p^6)$}}
\\
&=& (0.021\pm 0.003) \: \mbox{fm}^2
\end{eqnarray}
where the error is due to uncertainties in the ${\cal L}^{(4)}$-parameters
$L_i$ involved. This is to be compared with the experimental point 
$\langle r^2 \rangle_{\mbox{\em\scriptsize exp}} 
=-(0.025\pm 0.041)\:\mbox{fm}^2$ which is based on the data
\begin{eqnarray}
\langle r^2 \rangle^{\pi^+} &=& (0.439\pm 0.008) \: \mbox{fm}^2 
                                \quad \mbox{\cite{picexp}}
\\
\langle r^2 \rangle^{K^+} &=& (0.34\pm 0.05) \: \mbox{fm}^2
                                \quad \mbox{\cite{kacexp}}
\\
\langle r^2 \rangle^{K^0} &=& -(0.054\pm 0.026) \: \mbox{fm}^2
                                \quad \mbox{\cite{ka0exp}}
\\
\langle r^2 \rangle^{K\pi}_+ &=& (0.36\pm 0.02) \: \mbox{fm}^2
                                \quad \mbox{\cite{kp1exp}} \:.
\end{eqnarray}

\begin{figure} 
\begin{center}
\unitlength1mm
\begin{picture}(160,30) 
\put(72,5){\begin{picture}(0,0)\unitlength1mm\thinlines 
   \put(0,0){\line(-1,0){70}}
   \put(0,0){\vector(1,0){70}}
   \put(72,0){\makebox(0,0)[l]{$\displaystyle
      \frac{\langle r^2\rangle_{\mbox{\scriptsize\em Sirlin}}}{\mbox{fm}^2}$}}
   \put(60,-1){\line(0,1){2}}
   \put(60,-2){\makebox(0,0)[t]{0.03}}
   \put(40,-1){\line(0,1){2}}
   \put(40,-2){\makebox(0,0)[t]{0.02}}
   \put(20,-1){\line(0,1){2}}
   \put(20,-2){\makebox(0,0)[t]{0.01}}
   \put(0,-1){\line(0,1){2}}
   \put(0,-2){\makebox(0,0)[t]{0}}
   \put(-20,-1){\line(0,1){2}}
   \put(-20,-2){\makebox(0,0)[t]{-0.01}}
   \put(-40,-1){\line(0,1){2}}
   \put(-40,-2){\makebox(0,0)[t]{-0.02}}
   \put(-60,-1){\line(0,1){2}}
   \put(-60,-2){\makebox(0,0)[t]{-0.03}}
   \thicklines
   \put(-50,10){\circle*{2}}
   \put(-50,10){\line(1,0){82}}
   \put(-50,10){\line(-1,0){14}}
   \put(-70,10){\makebox(0,0)[l]{$\ldots$}}
   \put(-50,6){\makebox(0,0){\em experiment}}
   \put(0,20){\circle*{2}}
   \put(-9,24){\makebox(0,0){\em Sirlin's theorem}}
   \put(12,20){\circle*{2}}
   \put(15,16){\makebox(0,0){\em ChPT, $p^4$}}
   \put(42,20){\circle*{2}}
   \put(42,20){\line(1,0){6}}
   \put(42,20){\line(-1,0){6}}
   \put(42,16){\makebox(0,0){\em ChPT, $p^6$}}
   \end{picture}}
\end{picture}
\end{center}
\normalsize Figure 2:
\footnotesize The charge radius of Sirlin's linear combination, 
$\langle r^2\rangle_{\mbox{\tiny\em Sirlin}}
= \frac{1}{2} \langle r^2\rangle^{\pi^+} 
+ \frac{1}{2} \langle r^2\rangle^{K^+}
+ \langle r^2\rangle^{K^0} - \langle r^2\rangle^{K\pi}_+$:
Sirlin's theorem, the ${\cal O}(p^4)$ and ${\cal O}(p^6)$
predictions of ChPT and the experimental value.
\end{figure} 

We have actually calculated the complete $t$-dependence of $\Delta(t)$,
but we find only slight deviations from linearity \cite{PostSchilcher}.
From figure 2 we see that Sirlin's relation is poorly satisfied. 
Actually the ${\cal O}(p^6)$-contribution tends to increase the disagreement.
The main experimental uncertainty lies in the kaon form factors which
ought to be remeasured with higher accuracy. In particular it may be argued
\cite{GasserLeutw2} that the $K^+$ charge radius $\langle r^2\rangle^{K^+}$ 
should be larger than $\langle r^2 \rangle^{K\pi}_+$ which would bring 
prediction and experiment into better agreement.

We have checked our calculations in several ways:
\begin{itemize}
\item[-] 
In the special case of all masses equal, the irreducible two-loop 
integrals were compared to known analytic results \cite{BessisPusterla}.
\item[-]
The electromagnetic form factors have to satisfy the Ward identity.
This holds seperately for the group of reducible and the group of
irreducible diagrams.
\item[-]
Non-polynomial divergences have to disappear in the sum of all 
loop-diagrams.
\end{itemize}

In this note we have reported the results of the first two-loop or 
${\cal O}(p^6)$ calculation of a form factor in full chiral
$SU(3)\times SU(3)$ perturbation theory. We chose a particular
combination of weak and electromagnetic form factors due to Sirlin 
which is independent of the new arbitrary renormalization constants 
of ${\cal L}^{(6)}$ (except at $t=0$). The correction to the previous 
one-loop result \cite{GasserLeutw2} turns out to be significant. 
Comparison of Sirlin's linear combination of charge radii 
with data is inconclusive due to large experimental uncertainties
of the kaon charge radii. A significant improvement in the precision of
the charged kaon form factor should be feasible in the future COMPASS
experiment \cite{vHarrach}.

P.P. was supported by the ''Studienstiftung des deutschen Volkes''.


\end{document}